\documentclass[10pt,twocolumn]{opticajnl}
\journal{opticajournal} % use for journal or Optica Open submissions

\setboolean{shortarticle}{true}
\usepackage[toc]{appendix}
\usepackage{xr}
\usepackage{lineno}
\usepackage{graphicx} 
\usepackage{array}
\usepackage{hyperref,amsmath,amsthm,amssymb, multicol, array, bbm, enumitem, physics}
\usepackage{cleveref}
\usepackage{lineno}
\usepackage{titlesec}
\usepackage{comment}

\newcommand{\bs}{\mathbf{s}}
\newcommand{\bx}{\mathbf{x}}
\newcommand{\bk}{\mathbf{k}}

\newcommand{\by}{\mathbf{y}}

\newcommand{\hp}{{\phi}}

\newcommand{\hs}{{\sigma}}

\usepackage{float}
\usepackage{longtable}
\usepackage{physics}
\usepackage{ulem}

\title{One-photon communication in atomic media}

\author[1]{Zixiang Hong}
\author[2,3,*]{John C. Schotland}

\affil[2]{Department of Applied and Computational Mathematics, Yale University, New Haven, Connecticut 06511, USA}
\affil[2]{Department of Mathematics, Yale University, New Haven, Connecticut 06511, USA}
\affil[3]{Department of Physics, Yale University, New Haven, Connecticut 06511, USA}
\affil[*]{Corresponding author: john.schotland@yale.edu}

\begin{abstract}
We develop a framework to model single-photon transmission through an atomic medium, using quantum channel fidelity to describe the resulting information loss. We 
find that the normalized fidelity decreases monotonically with coupling strength, establishing a performance bound for quantum communication through such media. Our results hold for several channel types and for deterministic and random media.

\end{abstract}

\setboolean{displaycopyright}{false} 

\begin{document}

\maketitle

Quantum communications with single photons is a topic of fundamental interest and considerable applied importance \cite{Nielsen_Chuang_2010,RevModPhys.74.145}.  Various protocols, including quantum key distribution \cite{BEN84,Ekert91}, have been intensively studied and demonstrated in landmark experiments \cite{Rosenberg07,Liao18,Yin16}. Nevertheless, a significant practical challenge stems from the unavoidable interaction of photons with the physical medium through which they propagate \cite{Staudt07}. These interactions are a primary source of information loss, yet they are often accounted for phenomenologically using master equation approaches \cite{BRE02,Mele24,Cerf05,Leviant:2022uqm, lau_high-fidelity_2019,ZHAO07} and random matrix theory \cite{Benet11}, or neglected entirely in simplified theoretical treatments \cite{Rohde12}.  More sophisticated analyses, such as the fidelity calculation in \cite{Zhang10}, are limited by the assumptions of a medium consisting of a single atom. A comprehensive understanding of such effects is crucial for accurately predicting and  improving the performance of practical quantum communication systems.

In this Letter, we investigate the influence of an atomic medium on
single-photon communications. Our approach is based on the quantum field theory of a scalar field interacting with a system of two-level atoms. In order to characterize the information loss arising from atom-field interactions, we consider three quantum channels—erasure, completely dephasing, and depolarization. Our analysis demonstrates that the dependence of the fidelity on the coupling strength is described by a simple formula. We find that the fidelity, suitably normalized, is identical for the erasure and completely dephasing channels, and for uniform and random arrangements of atoms. The fidelity decreases monotonically as the coupling strength increases, indicating that the atom-field interaction has a detrimental effect on single-photon communication. Our results establish that, despite differences in the underlying physical mechanism, the degradation of quantum information follows a universal principle. We also identify the specific regime in which this principle breaks down for the depolarization channel.

% we emphasize that we utilize a first-principles theory; the environment is accounted for from first principles, in contrast to theories which add noise in a phenomenological manner.

We begin by considering the following model of a quantized field interacting with a collection of two-level atoms. For simplicity, we ignore the effects of polarization and employ a scalar theory of the electromagnetic field. Following the real-space quantization formalism presented in \cite{Kraisler2022}, the total Hamiltonian of the system is of the form
$
 {H} = {H}_F+ {H}_A + {H}_I. 
$
 The Hamiltonian of the field is given by
\begin{equation}
{H}_F = \hbar c \int d^3 x (-\Delta)^{1/2} \hp^\dagger(\bx)\hp(\bx),
\end{equation}
where $c$ is the speed of light and the operator $\hp^\dagger(\bx)$ ($\hp(\bx)$) creates (annihilates) a photon at the point $\bx$ and satisfies the commutation relations
$
[\hp(\bx),\hp^\dagger(\bx')] = \delta(\bx-\bx') \ \text{and} \ [\hp(\bx),\hp(\bx')] = 0.
$
The Hamiltonian of the atoms is given by
\begin{equation}
{H}_A = \hbar \Omega \int d^3 x n(\bx) \hs^\dagger(\bx) \hs(\bx),
\end{equation}
where $n$ is the number density of atoms and $\Omega$ is the atomic resonance frequency. The atomic operator $\hs^\dagger(\bx)$ ($\hs(\bx)$) creates (destroys) an excitation at $\bx$ and obeys the commutation relations
$
[\hs(\bx),\hs^\dagger(\bx')] = \frac{1}{n(\bx)} \delta(\bx - \bx') \ \text{and} \ [\hs(\bx),\hs(\bx')] = 0.
$
The atom-field interaction is described by the Hamiltonian
\begin{equation}
{H}_I = \hbar g \int d^3 x n(\bx)(\hp^\dagger(\bx)\hs(\bx) + \hp(\bx)\hs^\dagger(\bx)),
\end{equation}
where $g$ denotes the coupling strength and we have imposed the rotating wave approximation.
We suppose the system is in a one-excitation state of the form
\begin{equation}
\label{eq:SinglePhtonState}
\ket{\Psi} = \int d^3 x [\psi(\bx) \hp^\dagger(\bx) + n(\bx) a(\bx) \hs^\dagger(\bx)] \ket{\varnothing},
\end{equation}
where $\ket{\varnothing}$ is the combined vacuum state of the field and the ground state of the atoms. Here $\psi(\bx)$ and $a(\bx)$ are the probability amplitudes for finding a photon or an excited atom at $\bx$, respectively. The state $\ket{\Psi}$ obeys the the time-independent Schrodinger equation ${H}\ket{\Psi} = \hbar\omega \ket{\Psi}$, where $\hbar \omega$ is the energy of the system. It follows that $\psi$ and $a$ obey  
\begin{align}
c(-\Delta)^{1/2} \psi + gn(\bx) a &= \omega \psi , \\
g \psi + \Omega a &= \omega a. 
\end{align}
By eliminating $a$ from the above system, we find that $\psi$ obeys the equation
\begin{equation}
\label{eq:Main_Equation}
(-\Delta)^{1/2} \psi  + \frac{g^2 n(\bx)}{c(\omega - \Omega)} \psi = \frac{\omega}{c} \psi.
\end{equation}
The state $\ket{\Psi}$ obeys the normalization condition 
\begin{equation}
\begin{split}
\label{eq:NormalizationCondition}
\int d^3 x (\abs{\psi(\bx)}^2 + n(\bx)\abs{a(\bx)}^2) = 1,
\end{split}
\end{equation}
which guarantees the conservation of probability.

%influence of the atom-field interaction on the 

We now consider the transmission of quantum information by a channel. Since we are interested in channels that transmit photons, we treat the atoms as an unobserved component of the system. Therefore, we trace out the atomic degrees of freedom in the density matrix  $\rho(g)=\ket{\Psi(g)}\!\bra{\Psi(g)}$, where the dependence of $\ket{\Psi}$ on the coupling $g$ has been made explicit. We thereby obtain 
the following reduced density matrix for the photon amplitude
\begin{equation}
\begin{split}
\sigma(g) &= \Tr_{A} \rho(g) \\
&=\int d^3 x d^3 y \psi(\bx,g) \psi^\ast(\by,g) \hp^\dagger(\bx) \ket{0}\!\bra{0} \hp(\by) \\
&+ \int d^3  x n(\bx)\abs{a(\bx,g)}^2 \ket{0}\!\bra{0},
\end{split}
\end{equation}
where $\ket{0}$ represents the vacuum state of the field. The normalization condition  \eqref{eq:NormalizationCondition} ensures that $\Tr \sigma(g) = 1$ for all $g$. When $g=0$, the system is completely decoupled from the atomic medium and $\sigma(0) = \ket{\Phi}\!\bra{\Phi}$, where
\begin{equation}
\ket{\Phi} = \int d^3x\psi(\bx,g=0) \hp^\dagger(\bx) \ket{0} .
\end{equation}
The pure state $\ket{\Phi}$ is the input to the channel prior to transmission.

A quantum channel is a completely positive and trace-preserving operator that maps density matrices to density matrices.
%acts on the density matrix $\sigma(g)$. 
The fidelity $F_\mathcal{N}(g)$ of the channel $\mathcal{N}$ is defined by
\begin{equation}
\begin{split}
\label{eq:FormulaForFidelity}
F_\mathcal{N}(g)
&= \bra{\Phi} \mathcal{N}(\sigma(g))\ket{\Phi}.
\end{split}
\end{equation}
%\begin{comment}
% &= \left(\Tr\sqrt{\sqrt{\sigma(0)}\mathcal{N}(\sigma(g)) \sqrt{\sigma(0)}}\right)^2\\
%\end{comment}
We note that $F_\mathcal{N}(g)$ is a measure of the similarity of the input state $\sigma(0)$ is to the output state $\mathcal{N}(\sigma(g))$. A high value of $F_\mathcal{N}$ indicates that quantum information is relatively well preserved after accounting for decoherence due to interaction with the atomic medium and losses due to the channel. In our analysis, we compare the performance of three common channels. The erasure channel $\mathcal{N}_{E,p}$ with transmission rate $p \in [0,1]$ is defined as
\begin{equation}
\mathcal{N}_{E,p}(\sigma(g)) = p \sigma(g) + (1-p) \ket{0}\!\bra{0}.
\end{equation}
Here a photon is  transmitted through the channel with probability $p$ and lost with probability $1-p$. It follows from  \eqref{eq:FormulaForFidelity} that the fidelity of the erasure channel is given by
\begin{equation}
\label{eq:FidelityForErasureChannel}
F_{\mathcal{N}_{E,p}}(g) = p \abs{\int d^3 x \psi^\ast(\bx,0)\psi(\bx,g)}^2.
\end{equation}
In the limit of perfect transmission, where $p = 1$, this expression coincides with the fidelity of the identity channel $\mathcal{N}(\sigma(g))= \sigma(g)$.

Next, we consider the completely dephasing channel $\mathcal{N}_C$, which is a generalization of the dephasing channel for a single bosonic mode. %WHY?
%Physically, this corresponds to a scenario where random, uncontrolled phase shifts are applied to the quantum state. The effect is complete destruction of the coherent superposition of basis states. %WHY? Ans: Quantum phase here can be understood as the off-diagonal element in the desnity operator. Incohenrence means that the all off-diagonal terms are zero. A 2 Dimension example is as follow. Suppose any state can be written as superposition of basis state $\ket{\psi} = \alpha \ket{0} + \beta \ket{1}$. As density matrix, it is $\ket{\psi}\bra{\psi} = \abs{\alpha}^2\ket{0}\bra{0} + \abs{\beta}^2\ket{1}\bra{1} + \alpha \beta^\ast \ket{0}\bra{1} + \alpha^\ast \beta \ket{1}\bra{0}$. Suppose $N$ is a quantum channel that destroys all off-diagonal term which gives $N(\ket{\psi}\bra{\psi}) = \abs{\alpha}^2\ket{0}\bra{0} + \abs{\beta}^2\ket{1}\bra{1}$. Then this state cannot be written as a pure state thus we said destruction of coherent superposition.
The completely dephasing channel is defined by
\begin{equation}
\begin{split}
\mathcal{N}_C(\sigma(g)) &= \frac{V}{(2\pi)^3}\int d^3 x \abs{ \psi(\bx,g)}^2 \hp^\dagger(\bx) \ket{0}\!\bra{0} \hp(\bx) \\
&+ \int d^3 x n(\bx)\abs{a(\bx,g)}^2 \ket{0}\!\bra{0},
\end{split}
\end{equation}
where $V$ is the volume of the system \cite{TracePreservingForCompletelyDephasingChannel}. We note that the effect of $\mathcal{N}_C$ is to remove the off-diagonal elements of the density matrix $\sigma(g)$, thereby introducing a decoherence mechanism. 
Making use of \eqref{eq:FormulaForFidelity}, we find that the fidelity is given by
\begin{equation}
\begin{split}
\label{eq:FidelityForCompletelyDephasingChannel}
F_{\mathcal{N}_C}(g)
&= \frac{V}{(2\pi)^3}\int d^3 x \abs{\psi(\bx,0)}^2 \abs{\psi(\bx,g)}^2.
\end{split}
\end{equation}

Finally, we consider the depolarization channel, which generalizes the concept of depolarization from a single qubit to a continuum of modes. Its action is defined by
\begin{equation}
\label{def:DepolarizationChannel}
\mathcal{N}_{D,p}(\sigma(g))= p \sigma(g) + \frac{1-p}{\alpha} I_\Lambda.
\end{equation}
Here $p \in [0,1]$ is the transmission rate and $I_\Lambda$ is the band-limited identity operator with band-limit ${2\pi}/{\Lambda}$ and trace $\alpha$.  Physically, this channel describes a process in which the input state is preserved with probability $p$ and, with probability $1-p$, is replaced by a maximally mixed state, which may be interpreted as a uniform classical distribution over frequency modes up to the band limit.
Using \eqref{eq:FormulaForFidelity}, we see that the fidelity is given by
\begin{equation}
\begin{split}
F_{ \mathcal{N}_{D,p}}(g) = p \abs{\int d^3 x \psi^\ast(\bx,0)\psi(\bx,g)}^2
+ \frac{1-p}{\alpha}.
\end{split}
\end{equation}
Evidently, the fidelity of the depolarization channel differs by a constant from that of the erasure channel. 

%\textcolor{red}{I'm not sure what the the depolarization channel adds. Is there another channel we could discuss?}

We now apply the above results to two different configurations of the atomic medium.
We begin with the case of a uniform medium with constant density $n_0$. To proceed, we note that \eqref{eq:Main_Equation} is a nonlinear eigenvalue problem with eigenfunctions
of the form
$
\psi(\bx) = A e^{i\bk\cdot\bx} ,
$
where $\bk$ is a fixed wavevector
and the eigenvalues
\begin{equation}
\omega(\bk, g) = \frac{\Omega + c \abs{\bk} - \sqrt{(\Omega - c \abs{\bk})^2 + 4g^2 n_0}}{2}
\end{equation}
satisfy the consistency condition $\omega(\bk,g=0) = c \abs{\bk}$, which holds in the absence of coupling to the atoms. The above eigenfunctions do not obey the normalization condition \eqref{eq:NormalizationCondition}. Thus we consider a Gaussian wavepacket of the form
\begin{equation}
\psi(\bx) = A {\int_{\abs{\bk} = \abs{\bk_0}}  e^{i \bk \cdot \bx} e^{-\frac{\abs{\bk - \bk_0}^2}{2 \sigma^2}}}d^3 k ,
\end{equation}
with center $\bk_0$, width $\sigma$ and amplitude $A$. 
We find that the fidelity for the erasure channel is given by
\begin{equation}
\label{eq:ConstantDensityErasureChannelFidelity}
F_{\mathcal{N}_{E,p}}(g) = p N(g),
\end{equation}
and the fidelity for the completely dephasing channel is given by
\begin{equation}
\label{eq:ConstantDensityDephasingChannelFidelity}
F_{\mathcal{N}_C}(g) = N(g) \frac{\lambda^5}{\sigma^5 (2\pi)^5} \frac{e^{-\frac{4\lambda^2}{\sigma^2}}}{(1- e^{-\frac{4\lambda^2}{\sigma^2}})^2} I_1^+(\sqrt{2}\lambda),
\end{equation} 
where the functions $N(g)$ and $I_n^{\pm}(\lambda)$ are defined as
\begin{equation}
N(g) = \frac{\abs{\omega(\bk_0,g) - \Omega}^2 }{\abs{\omega(\bk_0,g) - \Omega}^2 + g^2 n_0},
\end{equation}
and 
\begin{equation}
I_n^{\pm}(\lambda) = \int d\Hat{\bk} d\Hat{\bs} \frac{e^{\frac{\lambda^2}{\sigma^2} (\Hat{\bk} + \Hat{\bs})\cdot \Hat{\bk}_0} }{\abs{\Hat{\bk} \pm \Hat{\bs}}^n}.
\end{equation}

% \Omega \to \Omega + c|k| ? Ans: this expression has been removed
% why don't both roots contribute? They do in the time-dependent constant density problem. Ans: both roots corresponds to different eigenvalue problem. In principle, we can choose any eigenvalue. The reason why we choose the way it is now is because we want to compare initially (when g = 0), the photon is in the field. Otherwise, it will seems that the fidelity increases as the coupling strength increases. It actually makes sense since the initial state for that case corresponds to vacuum state \ket{0}. As the coupling strength increase the excitation will go to the photon field. If we want to use that other state, we need to compare with \ket{0} not the photonic state.
% Shouldn't we consider a linear combination of both roots that produces a normalizable state? Ans: in the new paper, it is now
% what is V? Ans: V is the volume of the whole space

Next we consider the case of a disordered medium. We assume the atomic density $n(\bx)$ is of the form
$
n(\bx) = n_0 (1+ \eta(\bx)),
$
where $n_0$ is a constant and $\eta(\bx)$ is a real-valued random field that accounts for statistical fluctuations in the density.  We further assume the correlations of $\eta$ are given by
$
\left<\eta(\bx)\right> = 0 \ \text{and} \
\left<\eta(\bx)\eta(\by)\right> = C(\bx-\by),
$
where $\left<\cdots\right>$ denotes statistical averaging. We also assume that the medium is statistically homogeneous and isotropic, meaning that the correlation function $C$ depends only on the quantity $\abs{\bx-\by}$.
The calculation of the average fidelity for the random medium hinges on the evaluation of the two-point correlation function $\left<\psi(\bx) \psi^\ast(\by)\right>$. We compute this quantity using the diffusion approximation to the radiative transport equation. For a white-noise disorder model, characterized by the correlation function $C(\bx) = C_0 \delta(\bx)$ where $C_0$ is a constant, we find that the average fidelity for the erasure channel is given by
\begin{equation}
\begin{split}
\label{eq:DiffusionApproxErasureFidelity}
\left<F_{\mathcal{N}_{E,p}}(g) \right> = \frac{p \lambda^2}{\sigma^2 4\pi} \frac{e^{-\frac{2\lambda^2}{\sigma^2}}}{(1 - e^{-\frac{4\lambda^2}{\sigma^2}})} N(g) I^-_2(\lambda),
\end{split}
\end{equation}
and the average fidelity for the completely dephasing channel in the diffusion approximation is given by
\begin{equation}
\begin{split}
\label{eq:DiffusionApproxDephasingFidelity}
\left<F_{\mathcal{N}_{C}}(g)\right>
&= \frac{\lambda^2}{\sigma^2(2\pi)^6} \frac{e^{-\frac{2\lambda^2}{\sigma^2}}}{(1 - e^{-\frac{4\lambda^2}{\sigma^2}})} N(g) I^-_2(\lambda).
\end{split}
\end{equation}
The details of the calculations are presented in the Supplemental Material.

The above results can be put into a common form. It will prove useful to
introduce the normalized fidelity $F_\mathcal{N}(g)/F_\mathcal{N}(0)$, thereby obtaining the  simple formula:
\begin{equation}
\label{eq:UniversalPrinciple}
\frac{F_\mathcal{N}(g)}{F_\mathcal{N}(0)} = \frac{\abs{\omega(\bk_0,g) - \Omega}^2 }{\abs{\omega(\bk_0,g) - \Omega}^2 + g^2 n_0}.
\end{equation}
We note that this result holds for the erasure and completely dephasing channels, and for both uniform and random media.
This finding suggests that a fundamental similarity governs the information loss, despite differences in the physical medium and the nature of the induced errors. We note that the universality arises because the quantum channels are, by construction, independent of the atom-photon coupling strength, which itself only alters the amplitude—not the phase—of the single-photon state. Consequently, the specific details of the physical setting and the quantum channel do not matter. To illustrate the above results, we set the dimensionless parameter ${c^3 n_0}/{\Omega^3}=1$ and $c\abs{\bk_0}/\Omega = \frac{1}{2}$ and $\bk_0$ is in the $x$-direction.
\begin{figure}[t]
\centering
\includegraphics[width=.8 \linewidth]{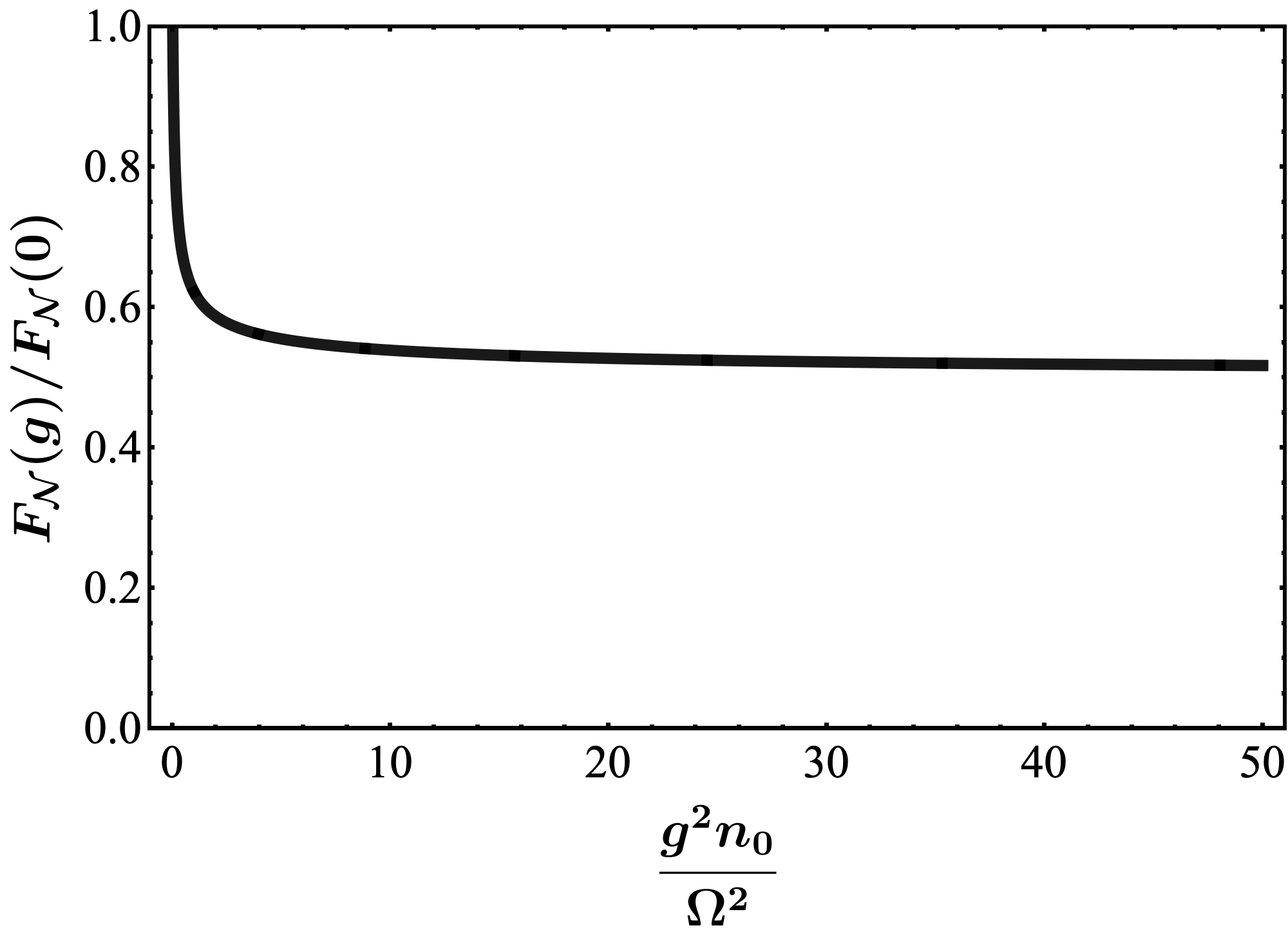}
\caption{Coupling-dependence of normalized fidelity.\label{fig:MainFigure}}
\end{figure} \\
As shown in \eqref{fig:MainFigure}, the normalized fidelity decreases monotonically with increasing coupling strength, yet it is fundamentally bounded from below. In the strong coupling regime ($g \to \infty$), the fidelity converges to a non-zero asymptote: 
\begin{equation}
\lim_{g \to \infty} \frac{F_\mathcal{N}(g)}{F_\mathcal{N}(0)}  = \frac{1}{2}.
\end{equation}
This result establishes a fundamental lower bound on fidelity, revealing that system performance does not degrade indefinitely with stronger coupling. It is worth noting that this asymptotic value of $1/2$ is analogous to the fidelity between a pure state and a maximally mixed state in a qubit system, suggesting a limit where information is substantially degraded but not completely erased.

To explore the boundaries of this principle, we examined its applicability to other decoherence models. We found that the depolarization channel, for instance, deviates from this universal behavior. For this channel in a constant density medium, the normalized fidelity is given by 
\begin{equation}
\frac{F_\mathcal{N}(g)}{F_\mathcal{N}(0)} = \left(p\frac{\abs{\omega_0(g)-\Omega}^2}{\abs{\omega_0(g)-\Omega}^2 +g^2 n_0} + \frac{1-p}{\alpha}\right)\frac{1}{p + \frac{1-p}{\alpha}}.
\end{equation} 
For a transmission rate $p \in (0,1)$, this expression clearly does not reduce to the simple principle found earlier. We note, however, that the simple principle is restored in the asymptotic limit of $\alpha \to \infty$, where the channel accommodates all frequency modes.

While fidelity provides a useful measure of state preservation, the channel capacity offers a more direct and fundamental quantification of information loss. We numerically calculated the capacity for the uniform density case by maximizing the Holevo information over a suitable ensemble of input states, as detailed in the Supplement. A key technical challenge is the infinite-dimensional nature of the continuous-variable system, which would otherwise permit an infinite capacity. To obtain a physically meaningful and finite result, we imposed an effective bandwidth constraint by discretizing the wave-vector space. For the erasure, completely dephasing, and depolarizing channels, the results show a clear and consistent trend: the capacity decreases monotonically as the coupling strength $g$ increases. This provides quantitative confirmation that the atom-field interaction is fundamentally detrimental to information transmission, acting as an intrinsic source of degradation regardless of the specific decoherence model.

In conclusion, we have investigated the effect of an atomic medium on single-photon quantum communication by developing a formalism from the first principles of a quantized electromagnetic field interacting with two-level atoms. To quantify the degradation of the quantum state—a source of information loss distinct from photon absorption—we introduced three quantum channels—erasure, completely dephasing, and depolarization—and derived their corresponding fidelities.

The central result of our study is the identification of a universal principle that governs fidelity for the erasure and completely dephasing channels. This law reveals a consistent and predictable behavior for these channels: for media with either uniform or random atomic densities, the normalized fidelity decreases monotonically with increasing coupling strength. Crucially, this decay does not lead to complete information loss. Instead, the normalized fidelity converges to an asymptote of $1/2$ in the strong coupling limit. This trend is corroborated by our calculations of the channel capacity, which also decreases monotonically with coupling strength for all three channels.  This set of findings establishes a direct, quantitative link between a fundamental physical interaction and the integrity of these quantum channels.

Several avenues for future research are apparent. A natural extension is to investigate two-photon systems to explore the medium's effect on entanglement. Furthermore, our framework can be applied to specific quantum communication protocols, such as quantum key distribution (QKD), to directly link channel fidelity with the secure key rate. Finally, a crucial practical challenge is to mitigate these detrimental effects by designing quantum error correction codes tailored to the specific characteristics of quantum channels.

\begin{backmatter}
\bmsection{Acknowledgments}
The authors would like to thank Anna C. Gilbert and Imran M. Mirza for helpful discussions.
\end{backmatter}

% Bibliography
\bibliography{sample}

\begin{thebibliography}{10}
\newcommand{\enquote}[1]{``#1''}

\bibitem{Nielsen_Chuang_2010}
M.~A. Nielsen and I.~L. Chuang, \emph{Quantum Computation and Quantum
  Information: 10th Anniversary Edition} (Cambridge University Press, 2010).

\bibitem{RevModPhys.74.145}
N.~Gisin, G.~Ribordy, W.~Tittel, and H.~Zbinden, {\protect\JournalTitle{Rev.
  Mod. Phys.}} \textbf{74}, 145 (2002).

\bibitem{BEN84}
C.~H. Bennett and G.~Brassard, \enquote{{Quantum cryptography: Public key
  distribution and coin tossing},} in \emph{Proceedings of IEEE International
  Conference on Computers, Systems, and Signal Processing,}  (India, 1984), p.
  175.

\bibitem{Ekert91}
A.~K. Ekert, {\protect\JournalTitle{Phys. Rev. Lett.}} \textbf{67}, 661 (1991).

\bibitem{Rosenberg07}
D.~Rosenberg, J.~W. Harrington, P.~R. Rice, \emph{et~al.},
  {\protect\JournalTitle{Phys. Rev. Lett.}} \textbf{98}, 010503 (2007).

\bibitem{Liao18}
S.-K. Liao, W.-Q. Cai, J.~Handsteiner, \emph{et~al.},
  {\protect\JournalTitle{Phys. Rev. Lett.}} \textbf{120}, 030501 (2018).

\bibitem{Yin16}
H.-L. Yin, T.-Y. Chen, Z.-W. Yu, \emph{et~al.}, {\protect\JournalTitle{Phys.
  Rev. Lett.}} \textbf{117}, 190501 (2016).

\bibitem{Staudt07}
M.~U. Staudt, S.~R. Hastings-Simon, M.~Nilsson, \emph{et~al.},
  {\protect\JournalTitle{Phys. Rev. Lett.}} \textbf{98}, 113601 (2007).

\bibitem{BRE02}
H.~P. Breuer and F.~Petruccione, \emph{The theory of open quantum systems}
  (Oxford University Press, Great Clarendon Street, 2002).

\bibitem{Mele24}
F.~A. Mele, F.~Salek, V.~Giovannetti, and L.~Lami, {\protect\JournalTitle{Phys.
  Rev. A}} \textbf{110}, 012460 (2024).

\bibitem{Cerf05}
N.~J. Cerf, J.~Clavareau, C.~Macchiavello, and J.~Roland,
  {\protect\JournalTitle{Phys. Rev. A}} \textbf{72}, 042330 (2005).

\bibitem{Leviant:2022uqm}
P.~Leviant, Q.~Xu, L.~Jiang, and S.~Rosenblum, {\protect\JournalTitle{Quantum}}
  \textbf{6}, 821 (2022).

\bibitem{lau_high-fidelity_2019}
H.-K. Lau and A.~A. Clerk, {\protect\JournalTitle{npj Quantum Information}}
  \textbf{5}, 31 (2019).

\bibitem{ZHAO07}
M.~ZHAO, T.~QIN, and Y.~ZHANG, {\protect\JournalTitle{Modern Physics Letters
  B}} \textbf{21}, 1531 (2007).

\bibitem{Benet11}
L.~Benet, S.~Hern\'andez-Quiroz, and T.~H. Seligman,
  {\protect\JournalTitle{Phys. Rev. E}} \textbf{83}, 056216 (2011).

\bibitem{Rohde12}
P.~P. Rohde, {\protect\JournalTitle{Phys. Rev. A}} \textbf{86}, 052321 (2012).

\bibitem{Zhang10}
Y.-Y. Zhang, Q.-H. Chen, and K.-L. Wang, {\protect\JournalTitle{Phys. Rev. B}}
  \textbf{81}, 121105 (2010).

\bibitem{Kraisler2022}
J.~Kraisler and J.~C. Schotland, {\protect\JournalTitle{Journal of Mathematical
  Physics}} \textbf{63}, 031901 (2022).

\bibitem{TracePreservingForCompletelyDephasingChannel}
{\protect\JournalTitle{When validating the trace}} -preserving property of the
  completely dephasing channel, an apparent divergence arises in the form of a
  $\delta(0)$ term. The divergence is resolved by considering the system within
  a finite volume, V. In a finite volume, the mode spectrum becomes discrete,
  which replaces the divergent Dirac delta function with a finite quantity.
  This standard regularization procedure confirms that the channel is indeed
  trace-preserving.

\end{thebibliography}

% Full bibliography added automatically for Optics Letters submissions; the following line will simply be ignored if submitting to other journals.
% Note that this extra page will not count against page length
\bibliographyfullrefs{sample}

\end{document}